\documentclass[]{interact}

\usepackage{slashbox}
\usepackage{multirow}
\usepackage{epstopdf}
\usepackage[caption=false]{subfig}
\usepackage[longnamesfirst,sort]{natbib}
\bibpunct[, ]{(}{)}{;}{a}{,}{,}
\renewcommand\bibfont{\fontsize{10}{12}\selectfont}
\theoremstyle{plain}

\theoremstyle{definition}

\theoremstyle{remark}

\begin{document}

%\articletype{ARTICLE TEMPLATE}% Specify the article type or omit as appropriate

\title{Analysing the efficiency of partially entangled states in Vaidman's-type games and its applications in Quantum Secret Sharing}

\author{
\name{Hargeet Kaur\textsuperscript{a}\thanks{CONTACT Hargeet Kaur. Email: kaur.1@iitj.ac.in} and Atul Kumar\textsuperscript{a}}
\affil{\textsuperscript{a}Indian Institute of Technology Jodhpur, Rajasthan}
}

\maketitle

\begin{abstract}
We analyse the role of degree of entanglement for Vaidman's game in a setting where the players share a set of partially entangled three-qubit states. Our results show that the entangled states combined with quantum strategies may not be always helpful in winning a game as opposed to the classical strategies. We further find the conditions under which quantum strategies are always helpful in achieving higher winning probability in the game in comparison to classical strategies. Moreover, we show that a special class of W states can always be used to win the game using quantum strategies irrespective of the degree of entanglement between the three qubits. Our analysis also helps us in comparing the Vaidman's game with the secret sharing protocol. Furthermore, we propose a new Vaidman-type game where the rule maker itself is entangled with the other two players and acts as a facilitator to share a secret key with the two players. For practical purposes, the analysis is extended to study the proposed game under noisy conditions. In addition, the results obtained here are also generalized for multi-qubit games.
\end{abstract}

\begin{keywords}
Vaidman's game; quantum secret sharing; entanglement; noise; GHZ and W
\end{keywords}

\section{Introduction}
Game theory is an eminently interesting and flourishing field of study, wherein many situations of conflicts can be efficiently examined and resolved \citep{GameTheory}. With the advent of quantum information and computation, game theory has generated a lot of interest in analysing quantum communication protocols from the perspective of a game \citep{BB84, BB84Game}. The analysis not only allows one to study the fundamental of quantum mechanics but also provides a much better insight to the communication protocol in terms of security, payoffs of different players, and complex nature of multi-qubit entanglement. The aim is to study and compare the payoffs of different users and security of a protocol using classical and quantum strategies. In general, quantum strategies are found to be preferable in comparison to the classical strategies. For example, Meyer demonstrated how quantum strategies can be utilized by a player to defeat his classical opponent in a classical penny flip game \citep{Meyer}. He further explained the relation of penny flip game setting to efficient quantum algorithms. Similarly, Eisert \citep{Eisert} suggested a solution based on quantum theory for avoiding the Prisoners' Dilemma. The quantum version of Prisoners' Dilemma game was also experimentally realized using a NMR quantum computer \citep{Du}. On the other hand, Anand and Benjamin \citep{Anand} found that for a scenario in penny flip game where two players share an entangled state, a player opting for a mixed strategy can still win against a player opting for a quantum strategy. Therefore, it becomes important to analyse the role of quantum entanglement in game theory. Furthermore, one must also understand and study the importance of using different entangled systems under different game scenarios to take the advantage of usefulness of such entangled systems in different situations.    \par
In this article, we analyse a game proposed by Vaidman \citep{Vaidman} in which a team of three players always wins the game, when they share a three qubit maximally entangled state. The team, however, does not win the game when players opt for pure classical strategies, in fact the maximum winning probability that can be achieved using classical strategies is $3/4$. Our analysis of Vaidman game includes two different classes of three-qubit entangled states, namely, $GHZ$ class \citep{GHZpaper} and $W$ class of states \citep{Dur}. We attempt to establish a             
relation between the winning probability of Vaidman's game \citep{Vaidman} with the degree of entanglement of various three-qubit entangled states used as a resources in the game. Interestingly, our results show that for $GHZ$ class, there are set of states for which classical strategies give better winning probability than the quantum strategies. In comparison to the $GHZ$ class, for a special class of $W$ states, quantum strategies prove to be always better than the classical strategies. We further establish a direct correspondence between  Vaidman's game and Quantum Secret Sharing (QSS) \citep{Hillery}. \par
In addition, we also propose a Vaidman-type game where one of the players sharing the three-qubit entanglement defines the rules of a game to be played between him/her and the other two players. A detailed examination of the proposed game shows that  the rule-maker finds himself in an advantageous situation whenever they share a partially entangled state, because this enables the rule-maker to modify rules in such a way that the team of other two players loose the game. We have further analysed the proposed game in a noisy environment, where we have considered that the qubits of the shared state may pass through an amplitude damping or a depolarizing channel or a phase flip channel.  Our results show that in case of W states, the winning probability using quantum strategy, still exceeds the classical winning probability for a phase flip channel. For GHZ state, the success probability using quantum strategy almost always exceeds the winning probability using classical strategies using both phase flip and amplitude damping channel. In all other cases, quantum strategies are found to be better than classical strategy for a certain range of decoherence parameters. Moreover, we further suggest an application of such a game in facilitated secret sharing between three parties, where one of the players is a facilitator and also controls the secret sharing protocol. In the later sections of the article, we have demonstrated the extension of Vaidman's game and the proposed game for multi-qubit scenario. Since the states used as resources in this article can be experimentally prepared \citep{Bouwmeester,Eibl,Dong}, the results obtained here may find applications in Quantum Secret Sharing or other similar protocols. \par
The organization of the article is as follows. In Section II and III, we briefly describe three-party entanglement and QSS, respectively. In section IV, we establish a correspondence of Vaidman's game with QSS, and in corresponding subsections we further demonstrate the outcomes of using $GHZ$ and $W$ class of states for Vaidman's game. A new Vaidman-type game is proposed in the Section V followed by its study in noisy conditions, and its application for QSS in the subsections. In section VI, a generalization of Vaidman's game for more than three players is discussed. To further extend the analysis, in section VII, the multi-player version of the game proposed in Section V is described. Finally, we conclude the article in Section VIII.

\section{Three-qubit Entanglement}
Dur et al. \citep{Dur} classified pure states of a three-qubit entangled systems in two inequivalent classes, namely $GHZ$ class and $W$ class represented as
%------------------------------------------------Eq. (1)-----------------------------------------------
\begin{equation}
\label{GHZgeneral}
\vert{\psi_{GHZ}}\rangle= sin\theta \vert{000}\rangle + cos\theta \vert{111}\rangle
\end{equation}
%-------------------------------------------------------------------------------------------------------
and
%-------------------------------------------------Eq. (2)-----------------------------------------------
\begin{equation}
\label{Wgeneral}
\vert{\psi_{W}}\rangle= a\vert{100}\rangle + b\vert{010}\rangle + c\vert{001}\rangle, 
\end{equation}
%-------------------------------------------------------------------------------------------------------
respectively where $\theta\in\left( 0,\pi/4\right)$ and $\left| a \right|^2  + \left| b \right|^2  + \left| c \right|^2  = 1$. The above two classes are termed as inequivalent classes as a state belonging to one of the class cannot be converted to a state belonging to another class by performing any number of Local Operations and Classical Communication (LOCC). The degree of entanglement for a pure three-qubit system can be defined using a measure called three-tangle $(\tau)$ \citep{Coffman}, given by 
%---------------------------------------------Eq. (3)--------------------------------------------------------
\begin{equation}
\label{tau_formula}
\tau= C^2_{P(QR)}-C^2_{PQ}-C^2_{PR}
\end{equation}
%------------------------------------------------------------------------------------------------------------
where $C_{P(QR)}$ represents the concurrence of the qubit P, with qubits Q and R taken together as one entity \citep{Hill, Wootters1, Wootters2}. The terms $C_{PQ}$ and $C_{PR}$ can be defined in a similar fashion such that,
%--------------------------------------------Eq. (4)-------------------------------------------------------
\begin{equation}
\label{conc_formula}
C(\vert{\psi}\rangle)=\vert\langle\psi\vert\sigma_y\otimes\sigma_y\vert\psi^*\rangle\vert
\end{equation}
%----------------------------------------------------------------------------------------------------------
Here, $\psi^*$ denotes the complex conjugate of the wave function representing the two-qubit entangled state. The value of three-tangle varies between 0 for product states to 1 for states having maximum entanglement. For example, the three-tangle for a state represented as $(a\vert{0}\rangle+b\vert{1}\rangle)\otimes(c\vert{00}\rangle+d\vert{11}\rangle)$ is 0 and for $GHZ$ states represented as   
%---------------------------------------------Eq. (5)------------------------------------------------------
\begin{equation}
\label{GHZ}
\vert{GHZ}\rangle= \frac{1}{\sqrt{2}}(\vert{000}\rangle+\vert{111}\rangle)
\end{equation}
%----------------------------------------------------------------------------------------------------------
is 1. Similarly, the standard state in $W$ class is represented by 
%----------------------------------------------Eq. (6)-----------------------------------------------------
\begin{equation}
\label{W}
\vert{W}\rangle= \frac{1}{\sqrt{3}}(\vert{001}\rangle+\vert{010}\rangle+\vert{001}\rangle)
\end{equation}
%----------------------------------------------------------------------------------------------------------
Although the standard $W$ state possesses genuine three-qubit entanglement, the same cannot be identified using the three-tangle as an entanglement measure as the three-tangle of the standard $W$ state is 0. Nevertheless, one can be assured that the $W$ class of states exhibit genuine tripartite entanglement using other entanglement measures such as average residual entanglement \citep{Dur} or sigma ($'\sigma'$) \citep{Emary}.

\section{Quantum Secret Sharing}
Secret sharing is the process of splitting a secret message into parts, such that no part of it is sufficient to retrieve the original message \citep{Hillery}. The original idea was to split the information between the two recipients, one of which may be a cheat (unknown to the sender). Only when the two recipients cooperate with each other, they retrieve the original message. The protocol, therefore, assumes that the honest recipient will not allow the dishonest recipient to cheat, hence, splitting the information between the two. \par
The original protocol can be implemented using the maximally entangled three-qubit $GHZ$ state, as given in (\ref{GHZ}), shared between three users Alice, Bob, and Charlie. Alice splits the original information between Bob and Charlie in a way that the complete message cannot be recovered unless they cooperate with each other. For sharing a joint key with Bob and Charlie, Alice suggests all of them to measure their qubits either in X or Y direction at random where the eigen states in X and Y basis are defined as
%-------------------------------------------------Eq. (7)-------------------------------------------------
\begin{equation}
\label{basis}
\vert{\pm x}\rangle=\frac{1}{\sqrt{2}}(\vert{0}\rangle \pm \vert{1}\rangle), \ \ \ \vert{\pm y}\rangle=\frac{1}{\sqrt{2}}(\vert{0}\rangle \pm i\vert{1}\rangle)
\end{equation}
%-------------------------------------------------------------------------------------------------------
\begin{table}[!t !h]
\renewcommand{\arraystretch}{1.3}
\caption{Effect of Bob's and Charlie's measurement on Alice's state in a $GHZ$ state}
\label{QSS}
\centering
\begin{tabular}{|c|cccc|}
\hline
\backslashbox{Bob}{Charlie} & $\vert{+x}\rangle$ & $\vert{-x}\rangle$ & $\vert{+y}\rangle$ & $\vert{-y}\rangle$ \\
\hline
$\vert{+x}\rangle$ & $\vert{+x}\rangle$ & $\vert{-x}\rangle$ & $\vert{-y}\rangle$ & $\vert{+y}\rangle$ \\
$\vert{-x}\rangle$ & $\vert{-x}\rangle$ & $\vert{+x}\rangle$ & $\vert{+y}\rangle$ & $\vert{-y}\rangle$ \\
$\vert{+y}\rangle$ & $\vert{-y}\rangle$ & $\vert{+y}\rangle$ & $\vert{-x}\rangle$ & $\vert{+x}\rangle$ \\
$\vert{-y}\rangle$ & $\vert{+y}\rangle$ & $\vert{-y}\rangle$ & $\vert{+x}\rangle$ & $\vert{-x}\rangle$ \\
\hline
\end{tabular}
\end{table}
The effects of Bob's and Charlie's measurements on the state of Alice's qubit is shown in Table \ref{QSS}. After performing their measurements at random, Bob and Charlie announce their choice of measurement bases (but not the measurement outcomes) to Alice. This is followed by Alice telling her choice of measurement basis to Bob and Charlie. Only the bases XXX, XYY, YXY, and YYX (for Alice, Bob, and Charlie, respectively) are accepted, for sharing the secret key. The results from all other random choices of bases are discarded. 

Bob and Charlie must meet and tell each other their measurement outcomes so as to collectively know the measurement outcome of Alice. For instance, if both Bob and Charlie measure in X basis and their measurement outcomes are $+1$($-1$) and $+1$($-1$) respectively, then the corresponding outcome of Alice will be $+1$ when measured in X basis. On the other hand, if the measurement outcomes of Bob and Charlie are $+1$ and $-1$ respectively or \textit{vice-versa}, then the corresponding outcome of Alice will be $-1$ when measured in X basis.

\section{Vaidman's Game representing Quantum Secret Sharing}
In this section, we show a correspondence between the QSS protocol \citep{Hillery} to the Vaidman's game \citep{Vaidman}. We, therefore, first briefly describe the Vaidman's game. In this game, three players, namely Alice, Bob, and Charlie, are taken to arbitrary remote locations: A, B, and C, respectively. Now each player is asked one of the two possible questions: Either \textquotedblleft{What is X?}\textquotedblright \ or \textquotedblleft{What is Y?}\textquotedblright . The players can give only two possible answers, either -1 or +1. The rules of the game suggest that either each player is asked the X question or two of the players are asked the Y question and the remaining one is asked the X question. The team of three players wins the game if the product of their answers is +1 (when all are asked the X question) and -1 (when one is asked the X question and two are asked the Y question). Clearly, if the players adopt the classical strategy then at best they can achieve a winning probability of $3/4$. On the other hand, if the three players share a three-qubit maximally entangled $GHZ$ state, as shown in (\ref{GHZ}), then they always win the game by using a simple quantum strategy, i.e., whenever a player is asked the X(Y) question, she/he measures her/his qubit in the X(Y) basis and uses the measurement outcome obtained in the measurement process as her/his answer.
\subsection{Use of GHZ class states in Vaidman's game}
That the three players always win the game using the above strategy is because of the strong correlations between the three qubits of the $GHZ$ state. For example, the three qubits in the $GHZ$ state are related as
%----------------------------------------------------Eq. (8)-----------------------------------------------
\begin{eqnarray}
\label{Vaidman Game}
\lbrace{M^X_A}\rbrace\lbrace{M^X_B}\rbrace\lbrace{M^X_C}\rbrace &=&1  \nonumber \\
\lbrace{M^X_A}\rbrace\lbrace{M^Y_B}\rbrace\lbrace{M^Y_C}\rbrace &=&-1 \nonumber \\
\lbrace{M^Y_A}\rbrace\lbrace{M^X_B}\rbrace\lbrace{M^Y_C}\rbrace &=&-1 \nonumber \\
\lbrace{M^Y_A}\rbrace\lbrace{M^Y_B}\rbrace\lbrace{M^X_C}\rbrace &=&-1 
\end{eqnarray}

%---------------------------------------------------------------------------------------------------------
\noindent where $\lbrace{M^X_i}\rbrace$ is the measurement outcome of the player $'i'$ measuring her/his qubit in X basis, and $\lbrace{M^Y_i}\rbrace$ is the measurement outcome of the player $'i'$ measuring her/his qubit in Y basis. A clear correspondence between the Vaidman's game and the QSS protocol is shown in (\ref{Vaidman Game}). \par

\begin{figure}[!h]
\centering
\resizebox*{8cm}{!}{\includegraphics{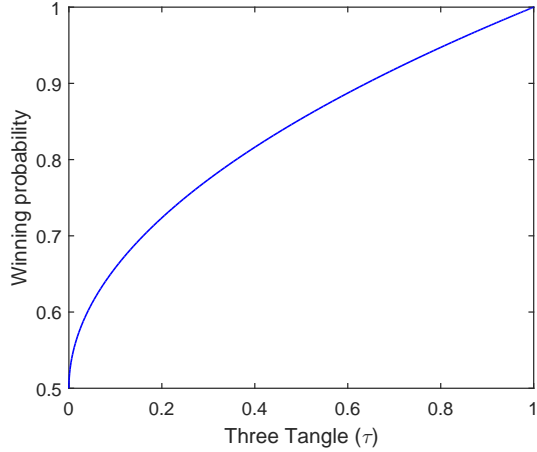}}\hspace{5pt}
\caption{Success probability of winning Vaidman's game using $GHZ$-type states}
\label{fig_genGHZ_VaidmanGame}
\end{figure}

We now proceed to analyse the Vaidman's game in a more general setting where the three players share a general $GHZ$ state represented in (\ref{GHZgeneral}), instead of sharing a maximally entangled $GHZ$ state as described in the original game. Clearly, for a general $GHZ$ state, the success probability of winning the above defined game varies from $0.5$ to $1$ as shown in Figure \ref{fig_genGHZ_VaidmanGame}. Here, we have assumed that the probability of players being asked the set of 4 questions ($XXX$, $XYY$, $YXY$, $YYX$) is equally likely. In Figure \ref{fig_genGHZ_VaidmanGame}, the winning probability of the game, i.e, $\frac{1}{2}(1+sin2\theta)$ is plotted against the degree of entanglement, three tangle ($\tau$). It is evident that only for maximally entangled state, i.e., when $\tau$ attains its maximum value (at $\theta=\pi/4$), the players have $100\%$ chances of winning the game. For all other values of $\tau$ the success probability is always less than the one obtained with a maximally entangled state. Interestingly, only the set of states with $\tau >0.25$ achieve better success probability in comparison to a situation where all the three players opt for classical strategies. Therefore, for the set of states with $\tau <0.25$, classical strategies will prove to be better in comparison to quantum strategies. Hence, for Vaidman's game, entanglement may not be always useful in winning the games using quantum strategies in comparison to classical strategies. 

\subsection{Use of W class states in Vaidman's game}
Although $W$-type states belong to a different class of states, they can also be used as resources in winning Vaidman's game with a different set of questions. In this case, the players may be asked either \textquotedblleft{What is Z?}\textquotedblright \ or \textquotedblleft{What is Y?}\textquotedblright. As earlier, the answers to these questions can again be either +1 or -1. For this, either all players are asked the Z question; or one of the players is asked the Z question and the remaining are asked the Y question. The players win the game if the product of their answers is -1, if all are asked the Z question; and +1, in all other cases. If the players share the standard $W$ state, given in (\ref{W}), before the start of play then they can win this game with a success probability of $0.875$. On similar grounds, we can use the standard $W$ state for probabilistic QSS, as QSS holds direct correspondence with the Vaidman's game. \par

\begin{figure}[!h]
\centering
\resizebox*{8cm}{!}{\includegraphics{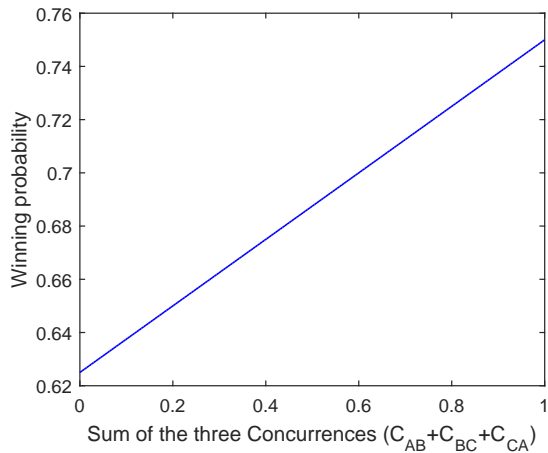}}\hspace{5pt}
\caption{Success probability of winning Vaidman's game using $W$-type states}
\label{fig_genW_VaidmanGame1}
\end{figure}

Similar to the case of $GHZ$ class, here, we analyse the success probability of the Vaidman's game if the three players share a general $W$-type state as shown in (\ref{Wgeneral}). In such a scenario, the team wins the game with a success probability of $\frac{1}{4}(\frac{5}{2}+bc+ab+ac)$. This value holds true for an assumption that the team will be asked the set of 4 questions ($ZZZ$, $ZYY$, $YZY$, $YYZ$) with equal probability. The plot of winning probability of Vaidman's game versus the sum of three residual concurrences is demonstrated in Figure \ref{fig_genW_VaidmanGame1}. The figure shows that the winning probability of Vaidman's game linearly increase with the sum of residual concurrences for $W$-type states. Furthermore, the plot also indicates that for $W$-type states, the winning probability of Vaidman's game is always greater than the classical winning probability if the sum of two qubit concurrences exceeds 1. Moreover, the highest success probability of $0.875$ can be achieved for $a=b=c=\frac{1}{\sqrt{3}}$. \par

\begin{figure}[!h]
\centering
\resizebox*{8cm}{!}{\includegraphics{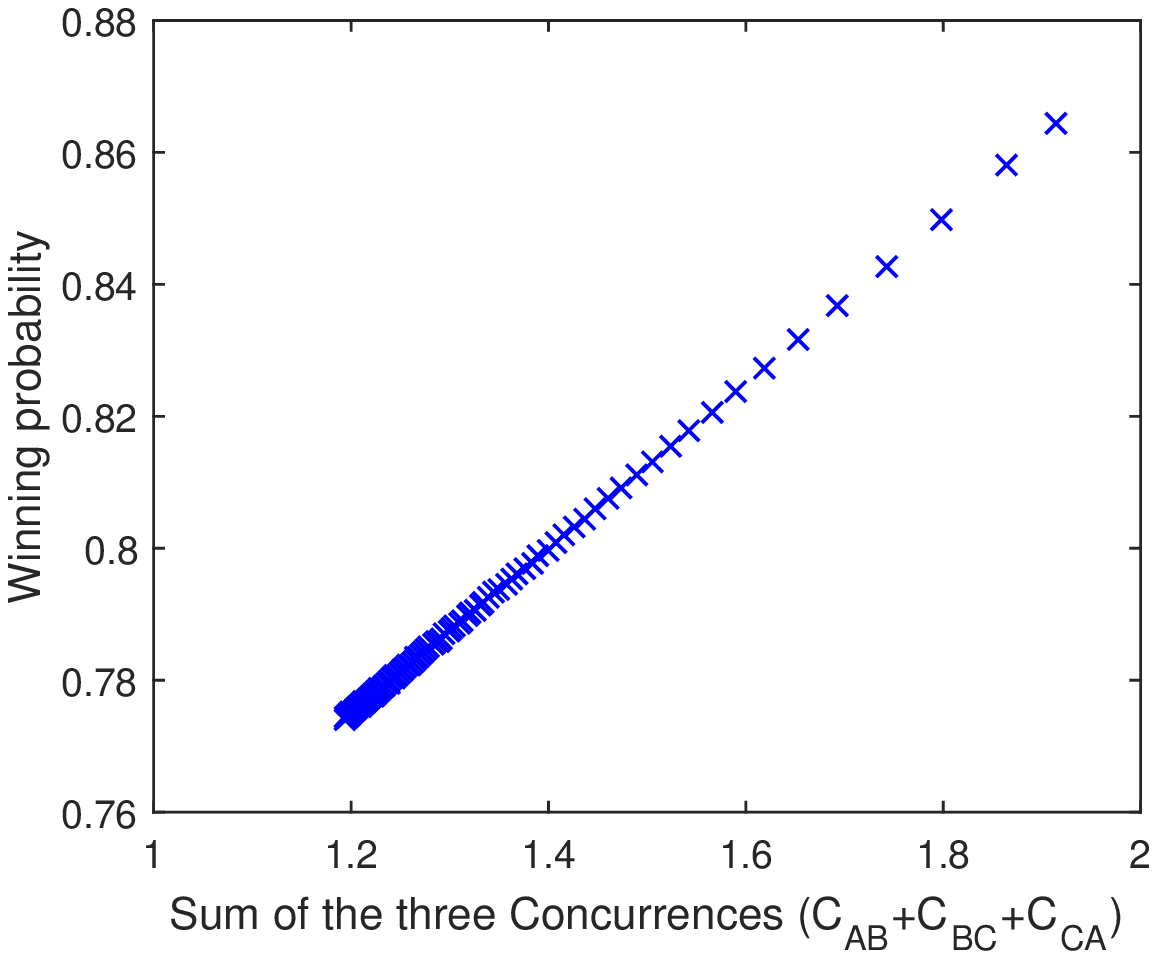}}\hspace{5pt}
\caption{Success probability of winning Vaidman's game using $W_n$ states}
\label{fig_Wn_VaidmanGame1}
\end{figure}

\begin{figure}[!h]
\centering
\resizebox*{8cm}{!}{\includegraphics{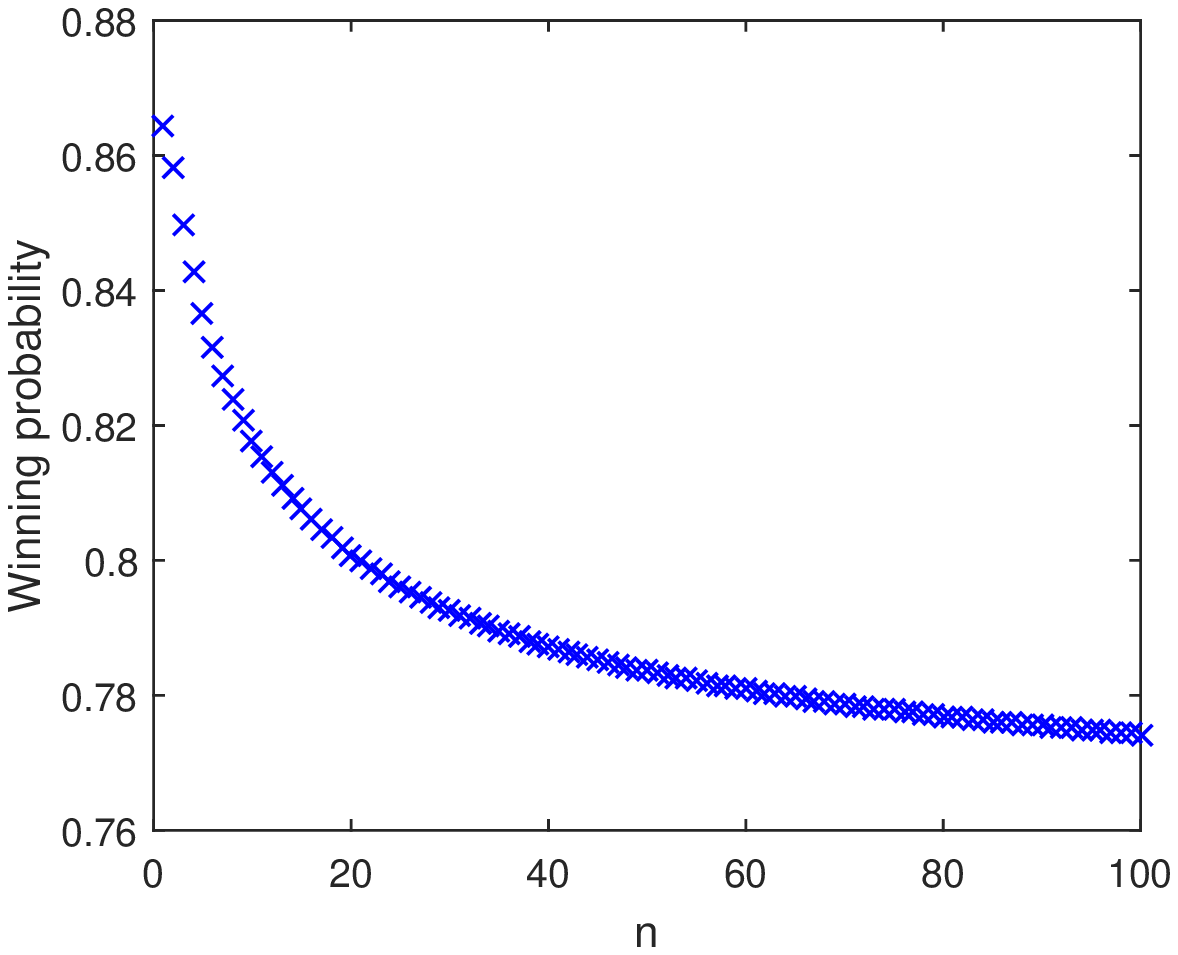}}\hspace{5pt}
\caption{Success probability of winning Vaidman's game using $W_n$ states}
\label{fig_Wn_VaidmanGame2}
\end{figure}

Although the use of partially entangled systems, in general, leads to probabilistic information transfer \citep{Karlsson, Shi}, Pati and Agrawal \citep{Agrawal} have shown that there exists a special class of $W$-type states which can be used for perfect teleportation and dense coding. Such states can be represented as 
%-----------------------------------------------Eq. (9)-----------------------------------------------------
\begin{equation}
\label{Wn}
\vert{W_n}\rangle= \frac{1}{\sqrt{2(1+n)}}(\vert{100}\rangle+\sqrt{n}e^{i\gamma}\vert{010}\rangle+\sqrt{n+1}e^{i\delta}\vert{001}\rangle)
\end{equation}
%------------------------------------------------------------------------------------------------------------

\noindent where $n$ is a positive integer and $\delta$ and $\gamma$ are relative phases. This motivates us to analyse the usefulness of these states for Vaidman's game. We found that the success probability of the game by sharing $W_n$ states as resources can be given by  $\frac{1}{8(n+1)}(5+5n+\sqrt{n+1}+\sqrt{n}(\sqrt{n+1}+1))$. Figure \ref{fig_Wn_VaidmanGame1} clearly depicts that if the three players share $W_n$ states, then the success probability using quantum strategies is always greater than the success probability using the classical strategies, independent of the value of sum of residual concurrences. Furthermore, Figure \ref{fig_Wn_VaidmanGame2} depicts the dependence of winning probability of Vaidman's game on parameter $n$. The highest success probability of $0.86425$ is achieved for $n=1$ when the sum of three residual concurrences is $1.914$. Nevertheless, the winning probability is always greater than the one obtained using classical strategies. 

\subsection{Comparison of the use of GHZ and W states}
The above analysis suggests that although a standard $GHZ$ state achieves $100\%$ success probability in winning the Vaidman's game which is more than the winning probability achieved by the standard $W$ state, only the set of $GHZ$-type states with a value of $\tau>0.25$ are useful for obtaining the success probability greater than the one obtained using classical strategies. Moreover, $W$-type states with the sum of three concurrences greater than one, can be useful in winning the game. In addition, a special class of $W$-type states, i.e. $W_n$ states give better prospects of winning the Vaidman's game, in comparison to any classical resource or strategy, for all values of $n$. 

\section{A two-player game where the rule-maker is entangled with the players}
The essence of Vaidman's game can be efficiently employed in an interesting scenario, where the rule-maker itself is entangled with the players playing the Vaidman-type game. In our proposed game, Alice, Bob and Charlie share a three-qubit entangled state. We assume that Charlie  prepares a three-qubit state and gives one qubit each to Alice ($A$) and Bob ($B$), keeping one ($C$) qubit with himself. Charlie agrees to help Alice and Bob, if they win the game as per the rules defined by Charlie. For this, Charlie measures his qubit in a general basis as shown in (\ref{Parametrized_basis}). Charlie, then asks questions \textquotedblleft{What is X?}\textquotedblright \ or \textquotedblleft{What is Z?}\textquotedblright \ to the team. Alice and Bob are not allowed to discuss and have to give individual answers each. Their answer can be +1 or -1.  If the team is asked the X (Z) question, both Alice and Bob measure their qubits in X (Z) basis and give their measurement results as answers to the asked questions. 
%-----------------------------------------------Eq. (10)-----------------------------------------------------
\begin{equation}
\label{Parametrized_basis}
\vert{b_0}\rangle= sin\lambda \vert{0}\rangle - cos\lambda \vert{1}\rangle; \ \ \ \ \
\vert{b_1}\rangle= cos\lambda \vert{0}\rangle + sin\lambda \vert{1}\rangle
\end{equation}

%-----------------------------------------------Eq. (11)-----------------------------------------------------
\begin{equation}
\label{rule1}
\vert{b_0}\rangle_C: \ \ \ \ \lbrace{M^X_A}\rbrace\lbrace{M^X_B}\rbrace=1 \ \ \ \ \ \lbrace{M^Z_A}\rbrace\lbrace{M^Z_B}\rbrace=-1
\end{equation}
%-----------------------------------------------Eq. (12)-----------------------------------------------------
\begin{equation}
\label{rule2}
\vert{b_1}\rangle_C: \ \ \ \ \lbrace{M^X_A}\rbrace\lbrace{M^X_B}\rbrace=-1 \ \ \ \ \ \lbrace{M^Z_A}\rbrace\lbrace{M^Z_B}\rbrace=1
\end{equation}

\begin{figure}
\centering
\resizebox*{8cm}{!}{\includegraphics{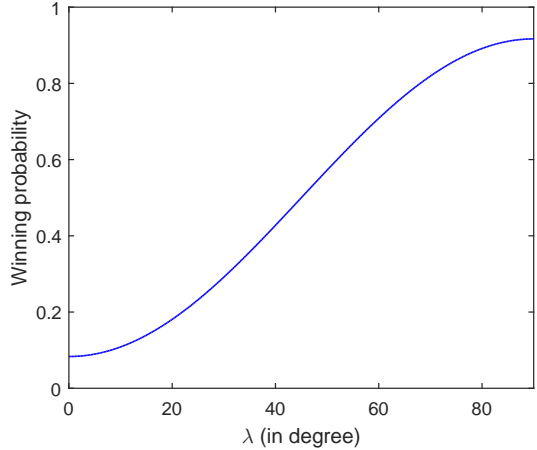}}\hspace{5pt}
\caption{Success probability of winning the proposed game where the rule-maker is entangled with the players using a standard $W$ state}
\label{fig_ProposedGame}
\end{figure}
If Charlie's measurement outcome is $\vert{b_0}\rangle$, he declares the winning condition to be the one listed in (\ref{rule1}), and if his  measurement outcome is $\vert{b_1}\rangle$, he declares the winning condition to be the one as listed in (\ref{rule2}). Here, $\lbrace{M^X_i}\rbrace$ is the measurement outcome when the player $'i'$ measures her/his qubit in X basis, and $\lbrace{M^Z_i}\rbrace$ is the measurement outcome when the player $'i'$ measures her/his qubit in Z basis.

We first consider that Charlie prepares a three-qubit $W$ state as shown in (\ref{W}). Clearly, the success probability of the team winning the game will depend on the parameter $\lambda$- governing the basis in which Charlie performs a measurement. It can be easily calculated that the value of the success probability achieved is $0.916667 - 0.833334 cos^2{\lambda}$. A plot of the success probability achieved with respect to the parameter $\lambda$ is shown in Figure \ref{fig_ProposedGame}. The maximum winning probability that the team can achieve is $0.9167$ for $\lambda=\dfrac{\pi}{2}$, i.e., when Charlie measures in computational basis ($\vert{b_0}\rangle=\vert{0}\rangle$ and $\vert{b_1}\rangle=\vert{1}\rangle$). On the other hand, if Charlie measures in computational basis $\vert{b_0}\rangle=\vert{1}\rangle$ and $\vert{b_1}\rangle=\vert{0}\rangle$, i.e. when $\lambda=0^\circ$, then the team mostly looses the game as the winning probability is only $0.0833$. Thus, if Charlie wants to help Alice and Bob, he prefers to prepare a standard $W$ state and performs measurement in the computational basis ($\vert{b_0}\rangle=\vert{0}\rangle$ and $\vert{b_1}\rangle=\vert{1}\rangle$) so that the team can win the game with a success rate of $91.667\%$. In this situation, the use of quantum strategy is always preferable for the team of Alice and Bob.  

If Charlie prepares a three-qubit $GHZ$ state as shown in (\ref{GHZ}) and shares it with Alice and Bob, then the team has only $50\%$ success probability irrespective of the measurement basis used by Charlie, which is equivalent to  a classical case where the team may choose not to measure its qubits and randomly answer as +1 or -1. However, if Charlie modifies the questions as X and Y and asks Alice and Bob to make a measurement in the X and Y basis, respectively, then in such a scenario the measurement outcome-dependent rules of the game would also be modified as:

%-----------------------------------------------Eq. (13)-----------------------------------------------------
\begin{equation}
\label{rule3}
\vert{b_0}\rangle_C: \ \ \ \ \lbrace{M^X_A}\rbrace\lbrace{M^X_B}\rbrace=-1 \ \ \ \ \ \lbrace{M^Y_A}\rbrace\lbrace{M^Y_B}\rbrace=+1
\end{equation}
%-----------------------------------------------Eq. (14)-----------------------------------------------------
\begin{equation}
\label{rule4}
\vert{b_1}\rangle_C: \ \ \ \ \lbrace{M^X_A}\rbrace\lbrace{M^X_B}\rbrace=+1 \ \ \ \ \ \lbrace{M^Y_A}\rbrace\lbrace{M^Y_B}\rbrace=-1
\end{equation}

\begin{figure}
\centering
\resizebox*{8cm}{!}{\includegraphics{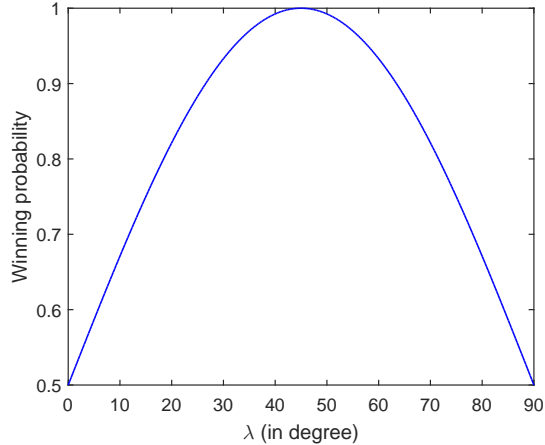}}\hspace{5pt}
\caption{Success probability of winning the proposed game where the rule-maker is entangled with the players using a standard $GHZ$ state}
\label{fig_ProposedGameGHZ}
\end{figure}

Therefore, if Charlie obtains $\vert{b_0}\rangle$ as his measurement outcome, then the outcomes of Alice and Bob satisfy (\ref{rule3}). On the contrary, if Charlie obtain $\vert{b_1}\rangle$ as his measurement outcome, then the outcomes of Alice and Bob satisfy (\ref{rule4}). The success chances of the team winning this game is $0.5(1+sin{2\lambda})$ and the maximum winning probability of 1 is attained for $\lambda=\dfrac{\pi}{4}$, i.e., when Charlie performs a measurement in diagonal basis ($\vert{-}\rangle$, $\vert{+}\rangle$). In general, Figure \ref{fig_ProposedGameGHZ} describes the success probability of the team as against the measurement parameter $\lambda$ when standard $GHZ$ state is used as a resource.

\subsection{Analysis of the proposed game in noisy environment}
In this subsection, we analyse the game discussed above in a noisy environment to study the nature and robustness of these states under real conditions and to study the effect of decoherence on the success probability of the proposed game. For this, we consider that Charlie prepares a three-qubit state and sends two qubits to Alice and Bob for the game to proceed. These two qubits may pass through a noisy channel, degrading the entanglement and correlation between qubits, and thus the success probability of the team (Alice and Bob) may also get affected. The quantum state $\rho$ after passing through a noisy channel changes to $\varepsilon(\rho)$ such that $\varepsilon(\rho)=\sum_{i} E_{i}{\rho}E_{i}^\dagger$ where $E_{i}s$ are the operation elements. The various types of noisy channels \citep{Nielsen} we consider  for our purpose are as follows:
\begin{itemize}
\item \textbf{Phase flip channel}: The phase flip channel flips the phase of the qubit ($\vert{0}\rangle$ to $\vert{0}\rangle$ and $\vert{1}\rangle$ to $-\vert{1}\rangle$) with probability $1-p$. The operation elements of this channel are:
$E_{0}=\sqrt{p}I=\sqrt{p}\left[ {\begin{array}{cc}
   1 & 0 \\
   0 & 1
    \\
  \end{array} } \right]
$ and
$E_{1}=\sqrt{1-p}Z=\sqrt{1-p}\left[ {\begin{array}{cc}
   1 & 0 \\
   0 & -1
    \\
  \end{array} } \right]
$
\item \textbf{Depolarizing channel}: When a qubit passes through a depolarizing channel, it gets depolarized to a completely mixed state $I/2$ with probability $p$. With probability $1-p$ the qubit is left untouched. The state of the quantum state $\rho$ after passing this channel is $\varepsilon(\rho)=p\dfrac{I}{2}+(1-p)\rho$
\item \textbf{Amplitude damping}: The operation elements of amplitude damping are $E_{0}=\left[ {\begin{array}{cc}
   1 & 0 \\
   0 & \sqrt{1-\gamma}
    \\
  \end{array} } \right]
$ and
$E_{1}=\left[ {\begin{array}{cc}
   0 & \sqrt{\gamma} \\
   0 & 0
    \\
  \end{array} } \right]
$
\end{itemize}

\begin{figure}[!h]
\centering
\resizebox*{9cm}{!}{\includegraphics{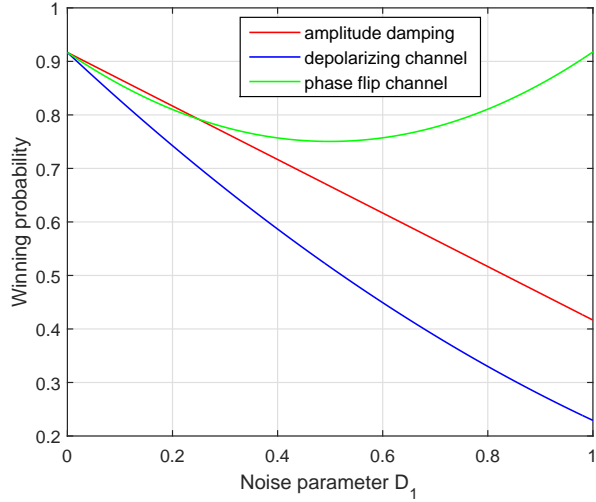}}\hspace{5pt}
\caption{Success probability of winning the game with respect to noise parameter ($D_1=D_2$) using the standard W state}
\label{noise_plot1}
\end{figure}

\begin{figure}[!h]
\centering
\resizebox*{9cm}{!}{\includegraphics{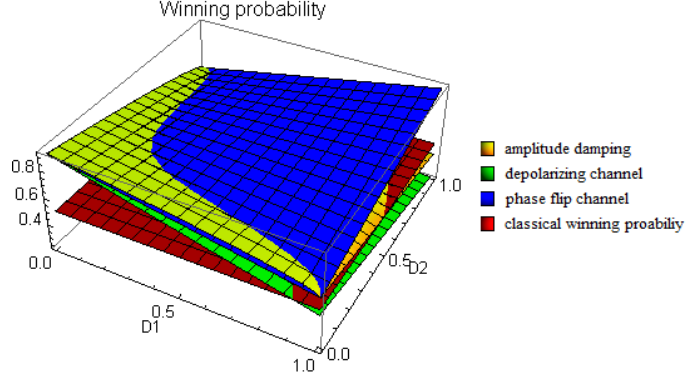}}\hspace{5pt}
\caption{Success probability of winning the game with respect to both the noise parameters ($D_1 \neq D_2$) using the standard W state}
\label{noise_plot2}
\end{figure}
 
In order to compare the effect of above noisy channels on the game using standard $W$ state as a resource, we evaluate the success probability of the game under noisy conditions. The success probability in such cases are listed in Appendix B. Figure \ref{noise_plot1} shows a plot of success probability of the game with respect to noise parameter $D_1$ (on qubit sent to Alice), assuming that the noise parameters on Alice's ($D_1$) and Bob's ($D_2$) qubit are equal. We further demonstrated a 3-D plot in Figure \ref{noise_plot2} showing variance between the winning probability of the game and the two noise parameters $D_1$ and $D_2$. Figure \ref{noise_plot1} and \ref{noise_plot2} clearly demonstrate that the game is most resistant to phase flip noise as the success rate of the game is always above the classical winning probability of 0.5. Moreover, our results show that the winning probability using the $W$ state is more robust towards the amplitude damping channel in comparison to the depolarizing channel. In both the cases however, the success probability falls below the classical winning probability, for higher values of noise parameters.

\begin{figure}[!h]
\centering
\resizebox*{9cm}{!}{\includegraphics{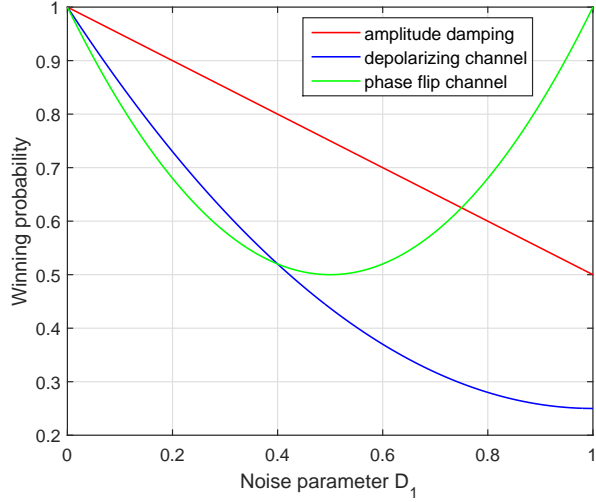}}\hspace{5pt}
\caption{Success probability of winning the game with respect to noise parameter ($D_1=D_2$) using a maximally entangled GHZ state}
\label{noise_plot1GHZ}
\end{figure}

\begin{figure}[!h]
\centering
\resizebox*{9cm}{!}{\includegraphics{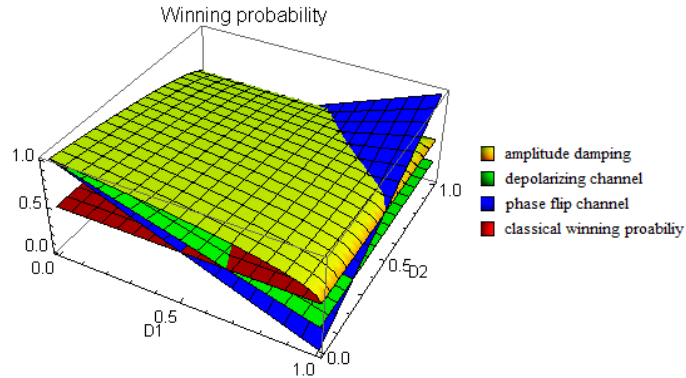}}\hspace{5pt}
\caption{Success probability of winning the game with respect to both the noise parameters ($D_1 \neq D_2$) using a maximally entangled GHZ state}
\label{noise_plot2GHZ}
\end{figure}

We have also evaluated the success probability of game when a $GHZ$ state is shared in a noisy environment. The results are depicted in Figure \ref{noise_plot1GHZ} and \ref{noise_plot2GHZ}. Figure \ref{noise_plot1GHZ} shows the relation between the winning probability of game and the noise parameter $D_1$ assuming that $D_1=D_2$. Further, Figure \ref{noise_plot2GHZ} describes the effect of both the parameters on success probability. These plots suggest that when both the noise parameters are equal, the game is resistant to phase flip as well as amplitude damping channel, as the success rate of the game is almost always greater than the classical winning probability of 0.5. However, in case of depolarizing channel, for high value of noise parameter, the winning probability falls below the classical case. Moreover, for $D_1 \neq D_2$, only the success probability in case of amplitude damping noise exceeds the classical winning probability. In other noisy environments, the winning probability may fall below the classical winning probability of $0.5$.

\subsection{An application of the above game in secret sharing}
For establishing a relation between the proposed game and secret sharing, we consider that Alice and Bob are kept in two different cells and are partially disallowed to communicate. By partially, we mean that they can communicate only under the presence of a facilitator or a controller (Charlie in our case), who listens and allows secure communication between the two. To accomplish this task, we prefer to exploit the properties of a standard $W$ state over the use of a $W_1$ state, because the success rate of winning Vaidman's game is $87.5\%$ when a standard $W$ state is shared, as opposed to $86.425\%$ when a $W_1$ state is shared within the team members. Also, we further consider that Charlie performs his measurement in the basis as shown in (\ref{Parametrized_basis}) at $\lambda=\dfrac{\pi}{2}$. \par

\begin{table}[!h]
\tbl{Control mode of facilitated information sharing}
{\begin{tabular}{c|cccccc} 
\hline
Charlie's measurement outcome & $\vert{1}\rangle$ & $\vert{1}\rangle$ & $\vert{1}\rangle$ & $\vert{1}\rangle$ & $\vert{1}\rangle$ & $\vert{1}\rangle$ \\
\hline
Alice's basis & Z & Z & X & X & X & X \\
Bob's basis & Z & X & Z & X & X & X \\
Is the choice of basis accepted? & yes & no & no & yes & yes & yes \\
Alice's measurement outcome & $+1$ & - & - & $+1$ & $-1$ & $+1$ \\
Bob's measurement outcome & $+1$ & - & - & $+1$ & $+1$ & $-1$ \\
\hline
Correlation as expected? & $\checkmark$ & - & - & $\times$ & $\checkmark$ & $\checkmark$ \\
\hline
\multicolumn{7}{c}{Alice and Bob are asked to announce their outcome and it is checked if} \\
\multicolumn{7}{c}{their results comply with (12) in more than or equal to $75\%$ cases} \\ 
\hline
\end{tabular}}
\label{table_CM}
\end{table}

\begin{table}[!h]
\tbl{Message mode of facilitated information sharing}
{\begin{tabular}{c|cccccc}
\hline
Charlie's measurement outcome & $\vert{0}\rangle$ & $\vert{0}\rangle$ & $\vert{0}\rangle$ & $\vert{0}\rangle$ & $\vert{0}\rangle$ & $\vert{0}\rangle$ \\
\hline
Alice's basis choice & X & X & X & Z & Z & Z \\
Bob's basis choice & X & X & Z & X & Z & Z \\
Basis choice accepted? & yes & yes & no & no & yes & yes \\
Alice's measurement outcome & $\vert{+}\rangle$ & $\vert{-}\rangle$ & - & - & $\vert{0}\rangle$ & $\vert{1}\rangle$ \\
Bob's measurement outcome & $\vert{+}\rangle$ & $\vert{-}\rangle$ & - & - & $\vert{1}\rangle$ & $\vert{0}\rangle$ \\
\hline
\multicolumn{7}{c}{$\vert{0}\rangle$ and $\vert{+}\rangle$ correspond to secret bit: 0} \\ 
\multicolumn{7}{c}{$\vert{1}\rangle$ and $\vert{-}\rangle$ correspond to secret bit: 1} \\
\hline
\multicolumn{7}{c}{Let Charlie announce that Bob should flip his outcome whenever he} \\ 
\multicolumn{7}{c}{chooses Z basis for measurement} \\
\hline
Shared secret bit & 0 & 1 & - & - & 0 & 1 \\
\hline
\end{tabular}}
\label{table_MM}
\end{table}

In order to share a key, Charlie chooses to operate in two different modes, namely control mode and message mode. The control mode corresponds to Charlie's measurement outcome $\vert{1}\rangle$, and is used to check whether Alice and Bob are honest or not, as shown in Table \ref{table_CM}. Similarly, the message mode corresponds to Charlie's measurement outcome $\vert{0}\rangle$, and is used to share a secret key with Alice and Bob (Table \ref{table_MM}). For this, Charlie prepares $'m'$ standard $W$ states as shown in (\ref{W}) and distributes qubits 1 and 2 of each state to Alice and Bob, respectively keeping the third qubit with himself. Charlie, then performs a measurement on his qubit in the computational ($\vert{0}\rangle$, $\vert{1}\rangle$) basis. Meanwhile, Alice and Bob randomly choose their bases of measurement (either X or Z) and announce their choice of bases to Charlie. If they choose two different bases, then their choices are discarded. Alternately, Charlie randomly chooses a basis of measurement and announces his choice to Alice and Bob. This will ensure that both Alice and Bob perform measurements in the same basis. This step is repeated for $'m'$ qubits, and Alice and Bob note down their measurement results each time. \par

If Charlie gets $\vert{0}\rangle$ as his measurement outcome, then he knows that the measurement results of Alice and Bob are related as in (\ref{rule1}) with certainty. As explained above, this will be the message mode of the proposed secret sharing scheme, wherein Alice's and Bob's outcomes will either be same or different. The relation between their outcomes is only known to Charlie, which he announces at the end of the protocol. On the other hand, if Charlie gets $\vert{1}\rangle$ as the measurement outcome, then the measurement results of Alice and Bob are related as in (\ref{rule2}) in $75\%$ cases. Since this is a control mode, Charlie secretly asks both Alice and Bob to announce their measurement outcomes, which he verifies to check if anyone (Alice or Bob) is cheating. If the results announced by Alice and Bob comply with the results in (\ref{rule2}) less than $75\%$ times, then cheating is suspected. Moreover, as Alice and Bob are not allowed to discuss, they cannot distinguish between the message and the control mode. If both, Alice and Bob are asked to announce their measurement outcomes, then the control mode of secret sharing is taking place. While, if none of them is asked to announce her/his results, then the message mode of secret sharing occurs. If Charlie suspects cheating in the control mode, he disallows communication and does not announce the relation between the outcomes of Alice and Bob for message runs. However, if Charlie does not find anything suspicious, he announces in the end, which results correspond to message and control mode, and also the relation between the outcomes of Alice's and Bob's measurement outcomes in the message mode. This protocol, therefore, enables the controller to check a pair of agents for their honesty, and simultaneous sharing of a secret key with them, if they are proved honest. 

Instead of sharing a $W$ state, if the players in the game share a $GHZ$ state, then Charlie performs his measurement in the diagonal basis as shown in (\ref{Parametrized_basis}) at $\lambda=\dfrac{\pi}{4}$. Here, the control mode corresponds to the measurement outcome $\vert{-}\rangle$ and the message mode corresponds to the measurement outcome $\vert{+}\rangle$. The protocol remains the same, i.e., the control mode is used to check the honesty of Alice and Bob and the message mode is used for sharing a mutual secret key between Alice and Bob. In this case, Alice and Bob randomly choose their bases of measurement (either X or Y) and announce their choice of bases to Charlie. If they choose two different bases, then their choices are discarded. 

If Charlie gets $\vert{+}\rangle$ as his measurement outcome, then he knows that the measurement results of Alice and Bob are related as in (\ref{rule4}) with certainty. This will be the message mode and the relation between the outcomes of Alice and Bob is only known to Charlie, which he announces at the end of the protocol. On the other hand, if Charlie gets $\vert{-}\rangle$ as the measurement outcome, then the measurement results of Alice and Bob are related as in (\ref{rule3}) in all cases. Similar to the previous protocol, Charlie secretly asks both Alice and Bob to announce their measurement outcomes. If the results announced by Alice and Bob do not always comply with the results in (\ref{rule3}), then cheating is suspected and the protocol is aborted, else it proceeds further so that the three players finally share a secret key, as in the case described above for the $W$ state. 

\section{Extension of Vaidman's game in higher dimensions}
For a three qubit system, Vaidman's game has 4 set of questions, XXX, XYY, YXY, and YYX, with answers +1, -1, -1, and -1  respectively. Similarly, for four, five, and six qubit systems, 8, 16, and 32 different types of questions, can be asked to the players in the game. For instance, if a four qubit state is shared between four players, then they can be asked the following 8 questions: XXXX, XXYY, XYXY, XYYX, YXXY, YXYX, YYXX, YYYY, i.e. all X questions, all Y questions, or two X and two Y questions. Depending on the set of questions that can be asked in a game, one can formulate games. For example, for a four-qubit case one can formulate a single game, but for a five or six qubits, one can formulate 2 or 3 distinct games respectively.

For more than three-player games, we realized that sharing a $W$ state between the players lead to the chances of win being less than the one that can be achieved classically. Therefore, with the increase in system's complexity and the number of players, $W$ states are not of much use for this type of game. The $GHZ$ states however are still useful and can be used as shared resources among the players, with a success probability of $100\%$ cases. Appendix A describes the rules of different four, five, and six player games and their winning condition when a $GHZ$ state is shared between the players. For example, in a 4-player game, either all players are asked the X question or two are asked X and two are asked Y question. The game is won if the product of the player's answers is -1 when all are asked question X, and if the product of the player's answers is +1 when two are asked question X and two are asked question Y. Classically the success probability of the game can not exceed $0.8517$. However, a four qubit $GHZ$ state with $\tau_4\geq0.51$ always gives a better winning probability than that achieved classically. Moreover, the players always win the game, when a maximally entangled $GHZ$ state is shared.

Similarly in a 5-player game, there are two possible scenarios. In the first one, either all players are asked question X, or two are asked question Y and three are asked question X. To win the game the team's answers must product to -1 in case of all X questions, and +1 in case of two Y and three X questions. Classically the maximum winning probability of the game is $0.909$. However, sharing any five qubit $GHZ$ class state with $\tau_5\geq0.67$ yields higher winning probability of the game, than by classical means. In another five-player scenario, either two players are asked Y question and remaining three are asked X question, or all except one player (who is asked X question) are asked Y question. Whenever two players are asked Y question, then product of the teams answers should be +1, and whenever four players are asked Y question, then product of the teams answers should be -1. Although, classically this game can be won with a success probability not more than $0.6667$, sharing a five-qubit $GHZ$ state with $\tau_5\geq0.11$, always leads to a winning probability greater than the classical one. Clearly, sharing the maximally entangled five-qubit GHZ state results in a $100\%$ win for the team. Appendix A further lists the outcomes of different 6 player games with GHZ states as resources.

\section{A three-player game where the rule maker is entangled with three players}
The following game is an extension of the one proposed in Section 5. In this game, Alice, Bob, Charlie, and Dave share a four-qubit entangled state. Dave prepares the four-qubit state and gives one qubit each to Alice ($A$), Bob ($B$), and Charlie ($C$), keeping one ($D$) qubit with himself. Dave is the rule-maker and thus decides the winning conditions for the team (Alice, Bob and Charlie). For this, Dave measures his qubit in a general basis as shown in (\ref{Parametrized_basis}), and then asks questions \textquotedblleft{What is X?}\textquotedblright \ or \textquotedblleft{What is Y?}\textquotedblright \ to the team. Alice, Bob and Charlie can individually give answer as +1 or -1 and are not allowed to discuss before answering. A player who is asked the X (Y) question, measures her/his qubits in X (Y) basis and gives her/his measurement result as the answer. 
%-----------------------------------------------Eq. (15)-----------------------------------------------------
\begin{eqnarray}
\label{4qubit_rule1}
\lbrace{M^X_A}\rbrace\lbrace{M^X_B}\rbrace\lbrace{M^X_B}\rbrace=+1 \ \ \ \ \ \lbrace{M^X_A}\rbrace\lbrace{M^Y_B}\rbrace\lbrace{M^Y_C}\rbrace=-1 \nonumber \\
\lbrace{M^Y_A}\rbrace\lbrace{M^X_B}\rbrace\lbrace{M^Y_C}\rbrace=-1 \ \ \ \ \ \lbrace{M^Y_A}\rbrace\lbrace{M^Y_B}\rbrace\lbrace{M^X_C}\rbrace=-1
\end{eqnarray}
%-----------------------------------------------Eq. (16)-----------------------------------------------------
\begin{eqnarray}
\label{4qubit_rule2}
\lbrace{M^X_A}\rbrace\lbrace{M^X_B}\rbrace\lbrace{M^X_B}\rbrace=-1 \ \ \ \ \ \lbrace{M^X_A}\rbrace\lbrace{M^Y_B}\rbrace\lbrace{M^Y_C}\rbrace=+1 \nonumber \\
\lbrace{M^Y_A}\rbrace\lbrace{M^X_B}\rbrace\lbrace{M^Y_C}\rbrace=+1 \ \ \ \ \ \lbrace{M^Y_A}\rbrace\lbrace{M^Y_B}\rbrace\lbrace{M^X_C}\rbrace=+1
\end{eqnarray}

\begin{figure}
\centering
\resizebox*{8cm}{!}{\includegraphics{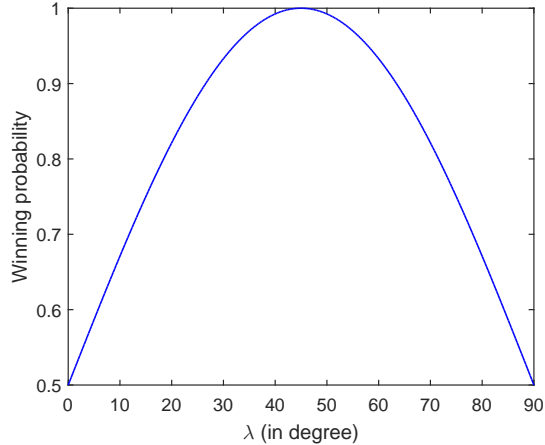}}\hspace{5pt}
\caption{Success probability of winning the proposed game where the rule-maker is entangled with the players using a 4 qubit maximally entangled $GHZ$ state}
\label{fig_4qubitProposedGame}
\end{figure}

If Dave's measurement outcome is $\vert{b_0}\rangle$, he declares the winning condition to be as shown in (\ref{4qubit_rule1}), and if his  measurement outcome is $\vert{b_1}\rangle$, he declares the winning condition to be as shown in (\ref{4qubit_rule2}). Here, $\lbrace{M^X_i}\rbrace$ is the measurement outcome when the player $'i'$ measures her/his qubit in X basis, and $\lbrace{M^Y_i}\rbrace$ is the measurement outcome when the player $'i'$ measures her/his qubit in Y basis.

%-----------------------------------------------Eq. (17)-----------------------------------------------------
\begin{equation}
\label{GHZ4}
\vert{GHZ}\rangle= \dfrac{1}{\sqrt{2}}(\vert{0000}\rangle-\vert{1111}\rangle)
\end{equation}

If Dave prepares a maximally entangled four-qubit state as shown in (\ref{GHZ4}), then the team wins the game with different winning probability for different values of parameter $\lambda$ (Figure \ref{fig_4qubitProposedGame}). Clearly $\lambda$ is a controlling parameter that controls the winning probability of the game for the other three players. From Figure \ref{fig_4qubitProposedGame}, we can observe that the maximum winning probability of 1 is achieved for $\lambda=\dfrac{\pi}{4}$, i.e., if Dave measures his qubit in diagonal basis {$\vert{-}\rangle$, $\vert{+}\rangle$}, the above game is always won by the players. Classically such a game can only be won in half the cases. Similar to the above proposed game, one can generalize different games in higher dimensions as well. This can also be extended for sharing a secret key among players in a similar manner as described in the subsection 5.2.

\section{Conclusion}
In this article, we addressed the role of degree of entanglement for Vaidman's game. We analysed the relation between the success probability of the Vaidman's game with the three-qubit entanglement measures considering both quantum and classical strategies. The results obtained here indicate that entanglement and quantum strategies may not be always useful in winning the game. For example, we found that there are set of $GHZ$ class and $W$ class states, for which classical strategies are proved to be better than the quantum strategies. On the other hand, for the special class of $W$-type states, i.e., $W_{n}$ states, quantum strategies are always better than the classical strategies in winning the Vaidman's game. We further explored a correspondence between the Vaidman's game using general three-qubit pure states and the QSS protocol. In addition, we have proposed an efficient game, where the player deciding the rules of the game itself is entangled with other two players. The proposed game may find an application in facilitated secret sharing, where a facilitator checks the players involved for their honesty and simultaneously controls the process of sharing information between them. \par 

We have also analysed these games under real situations, i.e., considering the success probability  of the game under noisy conditions, for example using amplitude damping, depolarizing channel and phase flip channel. Interestingly, it has been found that both $W$ and $GHZ$ states, when used as a shared quantum state in the game, are more robust to phase flip noise. Moreover, $GHZ$ states give better winning probability than that achieved classically, even when two of its qubits pass through an amplitude damping channel. Further, we have also extended our analysis for similar games between four, five and six players. In the game having more than three players, it has been found that $GHZ$ states are a useful resource for the proposed protocol, as they help attain $100\%$ winning probability. Furthermore, just like the three qubit proposed game holds application in secret sharing, the multi-qubit counterpart of the game, as discussed, will also hold similar utilization.

\newpage
\appendix
\section{Generalization of Vaidman's Game for Multi-qubit Systems}

\begin{table}[h!]
\renewcommand{\arraystretch}{1.3}
%\caption{Different extensions to Vaidman's game}
\label{tableA_extendedVG}
\centering
% Some packages, such as MDW tools, offer better commands for making tables
% than the plain LaTeX2e tabular which is used here.
\begin{tabular}{|c|c|c|c|}
\hline
Number & Winning conditions & Classical & Range of n-tangle \\
of & for the game & winning & $\tau_n$ of  $GHZ$ states \\
players && probability & for which quantum \\
&&& strategies exceeds \\
&&& classical strategy \\
\hline
4 & $XXXX=-1$ & $0.8517$ & $0.51\leq\tau_4\leq1$ \\
Game 1 & $XXYY=+1$ & & \\
& $XYXY=+1$ & & \\
& $XYYX=+1$ & & \\
& $YXXY=+1$ & & \\
& $YXYX=+1$ & & \\
& $YYXX=+1$ & & \\
\hline
5 & $XXXXX=-1$ & $0.909$ & $0.67\leq\tau_5\leq1$ \\
Game 1 & $YYXXX=YXYXX$ & & \\
& $=YXXYX=YXXXY$ & & \\
& $=XYYXX=XYXYX$ & & \\
& $=XYXXY=XXYYX$ & & \\
& $=XXYXY=XXXYY=+1$ & & \\
\hline
5 & $YYXXX=YXYXX$ & $0.6667$ & $0.11\leq\tau_5\leq1$ \\
Game 2 & $=YXXYX=YXXXY$ & & \\
& $=XYYXX=XYXYX$ & & \\
& $=XYXXY=XXYYX$ & & \\
& $=XXYXY=XXXYY=+1$ & & \\
& $XYYYY=YXYYY$ & & \\
& $=YYXYY=YYYXY$ & & \\
& $=YYYYX=-1$ & & \\
\hline
6 & $XXXXXX=-1$ & $0.9375$ & $0.765\leq\tau_6\leq1$ \\
Game 1 & $YYXXXX=YXYXXX$ & & \\
& $=YXXYXX=YXXXYX$ & & \\
& $=YXXXXY=XYYXXX$ & & \\
& $=XYXYXX=XYXXYX$ & & \\
& $=XYXXXY=XXYYXX$ & & \\
& $=XXYXYX=XXYXXY$ & & \\
& $=XXXYYX=XXXYXY$ & & \\
& $=XXXXYY=+1$ & & \\
\hline
\end{tabular}
\end{table}
%\newpage
\begin{table}[!t]
\renewcommand{\arraystretch}{1.3}
%\caption{Different extensions to Vaidman's game}
\label{tableB_extendedVG}
\centering
% Some packages, such as MDW tools, offer better commands for making tables
% than the plain LaTeX2e tabular which is used here.
\begin{tabular}{|c|c|c|c|}
\hline
Number & Winning conditions & Classical & Range of n-tangle \\
of & for the game & winning & $\tau_n$ of  $GHZ$ states \\
players && probability & for which quantum \\
&&& strategies exceeds \\
&&& classical strategy \\
\hline
6 & $YYXXXX=YXYXXX$ & $0.5$ & $0\leq\tau_6\leq1$ \\
Game 2 & $=YXXYXX=YXXXYX$ & & \\
& $=YXXXXY=XYYXXX$ & & \\
& $=XYXYXX=XYXXYX$ & & \\
& $=XYXXXY=XXYYXX$ & & \\
& $=XXYXYX=XXYXXY$ & & \\
& $=XXXYYX=XXXYXY$ & & \\
& $=XXXXYY=+1$ & & \\
& $XXYYYY=XYXYYY$ & & \\
& $=XYYXYY=XYYYXY$ & & \\
& $=XYYYYX=YXXYYY$ & & \\
& $=YXYXYY=YXYYXY$ & & \\
& $=YXYYYX=YYXXYY$ & & \\
& $=YYXYXY=YYXYYX$ & & \\
& $=YYYXXY=YYYXYX$ & & \\
& $=YYYYXX=-1$ & & \\
\hline
6 & $XXYYYY=XYXYYY$ & $0.9375$ & $0.765\leq\tau_6\leq1$ \\
Game 3 & $=XYYXYY=XYYYXY$ & & \\
& $=XYYYYX=YXXYYY$ & & \\
& $=YXYXYY=YXYYXY$ & & \\
& $=YXYYYX=YYXXYY$ & & \\
& $=YYXYXY=YYXYYX$ & & \\
& $=YYYXXY=YYYXYX$ & & \\
& $=YYYYXX=-1$ & & \\
& and $YYYYYY=+1$ & & \\
\hline
\end{tabular}
\end{table}

\
\newpage
\section{Winning probability of the three-qubit proposed game in a noisy environment}
\begin{table}[h!]
\renewcommand{\arraystretch}{1.3}
\label{tableB_noise}
\centering
\begin{tabular}{|c|c|c|}
\hline
Quantum State & Noise & Winning probability of the game \\
\hline
\multirow{ 6}{*}{W state} & Amplitude damping & $0.75-0.1667D_1-0.1667D_2$ \\
& & $+0.1667\sqrt{(1-D_1)(1-D_2)}$ \\
\cline{2-3}
& Depolarizing channel & $0.91667-0.45833D_1-0.45833D_2$ \\
& & $+0.229167D_1D_2$ \\
\cline{2-3}
& Phase flip channel & $0.91667-0.333D_1-0.333D_2$ \\
& & $+0.667D_1D_2$ \\
\hline
\multirow{ 3}{*}{GHZ state} & Amplitude damping &  $0.5 + 0.5\sqrt{(1-D_1)(1-D_2)}$ \\ 
\cline{2-3}
& Depolarizing channel & $1 - 0.75D_1 -0.75D_2 + 0.75D_1D_2$ \\
\cline{2-3}
& Phase flip channel & $1 - D_1 - D_2 + 2D_1D_2$ \\
\hline
\end{tabular}
\end{table}

\end{document}